\documentclass{article}
\usepackage{atbegshi}
\AtBeginDocument{\AtBeginShipoutNext{\AtBeginShipoutDiscard}}
\usepackage{url} 
\usepackage[a4paper,left=2.5cm,top=2.5cm,right=2.5cm,bottom=2.5cm]{geometry}
\usepackage{gensymb}
\usepackage{xcolor}
\usepackage{setspace}
\usepackage{varioref}
\usepackage{graphicx}        
\usepackage{rotating}
\usepackage{subcaption}
\usepackage{booktabs}
\usepackage{array}
\usepackage[verbose]{placeins}
\usepackage{amsmath}
\usepackage{rotating}
\usepackage[export]{adjustbox}
\usepackage{caption}
\usepackage{eucal}
\usepackage{amssymb}
\usepackage{mathrsfs}
\usepackage{amsmath}
\usepackage{float}
\restylefloat{table}
\usepackage{blindtext}
\usepackage{natbib}
\usepackage{multirow}
\usepackage{booktabs,caption,fixltx2e}
\usepackage[flushleft]{threeparttable}
\usepackage{tabularx}
\usepackage{bm}
\usepackage{marvosym} 
\usepackage{etoolbox}
\usepackage{tikz}
\newrobustcmd*{\mycircle}[1]{\tikz{\filldraw[draw=#1,fill=#1] (0,0) circle [radius=0.1cm];}}
\newrobustcmd*{\mytriangle}[1]{\tikz{\filldraw[draw=#1,fill=#1] (0,0) -- (0.2cm,0) -- (0.1cm,0.2cm);}}

\providecommand{\keywords}[1]{\textbf{\textit{Keywords---}} #1}
\begin{document}
 \pagenumbering{gobble} 

\title{A Bayesian Approach to Modelling Fine-Scale Spatial Dynamics of Non-State Terrorism: World Study, 2002-2013}
\author{Andr\'{e} Python\textsuperscript{1}, Janine Illian\textsuperscript{1}, Charlotte Jones-Todd\textsuperscript{1} and Marta Blangiardo\textsuperscript{2,3}}


\rule[0.5ex]{1cm}{0.4pt}

\maketitle
\pagenumbering{arabic}  
\footnotetext[1]{School of Mathematics and Statistics, University of St Andrews, UK}
\footnotetext[2]{Department of Epidemiology \& Biostatistics, Imperial College London, UK}
\footnotetext[3]{MRC-PHE Centre for Environment \& Health, Imperial College London, UK}
%
%
%


\begin{abstract}To this day, terrorism persists as a worldwide threat, as exemplified by the ongoing lethal attacks perpetrated by ISIS in Iraq, Syria, Al Qaeda in Yemen, and Boko Haram in Nigeria. In response, states deploy various counterterrorism policies, the costs of which could be reduced through efficient preventive measures. Statistical models able to account for complex spatio-temporal dependencies have not yet been applied, despite their potential for providing guidance to explain and prevent terrorism. In an effort to address this shortcoming, we employ hierarchical models in a Bayesian context, where the spatial random field is represented by a stochastic partial differential equation. Our results confirm the contagious nature of the lethality of terrorism and the number of lethal terrorist attacks in both space and time. Moreover, the frequency of lethal attacks tends to be higher in more economically developed areas, close to large cities, and within democratic countries. In contrast, attacks are more likely to be lethal far away from large cities, at higher altitudes, in less economically developed areas, and in locations with higher ethnic diversity. We argue that, on a local scale, the lethality of terrorism and the frequency of lethal attacks are driven by antagonistic mechanisms.
\end{abstract}
\vspace{1em}
\keywords{Bayesian hierarchical models; GMRF; space-time models; SPDE; terrorism}

\section{Introduction}
\label{sec:intro}
Nowadays, terrorism represents a worldwide threat, illustrated by the ongoing deadly attacks perpetrated by the Islamic State (also called ISIS, ISIL, or Daesh) in Iraq and Syria, Boko Haram in Nigeria, and the al-Nusra Front (also called Jabhat al-Nusra) in Syria for example. In response to this threat, states may use a combination of counterterrorism policies, which include the use of criminal justice, military power, intelligence, psychological operations, and preventives measures \citep[p.~45]{Crelinsten2009}. In particular, following the tragic terrorist attacks on September 11, 2001 (9/11) in New York, states have tended to increase their spending to counter terrorism. From 2001 to 2008, the expenditure of worldwide homeland security increased by US\$\,70 billion \citep{Nato2008}. In the US alone, the 2013 federal budget devoted to combat terrorism reached around US\$\,17.2 billion \citep{Washington2013}. 

Theoretical work and empirical studies at country-level pointed out that the causes of terrorism are complex and multidimensional and include economic, political, social, cultural, and environmental factors (\citealp{Brynjar2000}; \citealp[p.~60]{Richardson2006}; \citealp{Gassebner2011}; \citealp{Krieger2011}; \citealp{Hsiang2013}). Moreover, the relationship between specific factors and terrorism is often not straightforward. For example, \citet[p.~194]{Hoffman2006} described the role of media in covering terrorism as a ``double-edged sword'': publicity promotes terrorist groups, which facilitates the recruitment of new members and strengthens the cohesion of groups, but in turn, encourages society to marshal resources to combat terrorism \citep{Rapoport1996}
. At the individual level, the actions and beliefs of each member of terrorist groups are important drivers of terrorism as well (\citealp[p.29]{Crenshaw1983}; \citealp[p.141]{Wilkinson1990}; \citealp[pp.92-93]{Richardson2006}).

From a modelling perspective, terrorism is therefore not an entirely random process. Similar to contagious diseases \citep{Jacquez1996, Mantel1967, Loftin1986} or seismic activity \citep{Mohler2011,Crough1980,Courtillot2003}, terrorist events are rarely homogeneously distributed in space. In contrast, they tend to exhibit high concentration levels in specific locations (so-called ``hot-spots'') \citep{LaFree2009b,LaFree2010,LaFree2012, Nacos2010, Steen2006, Piegorsch2007}, and can spread from one location to another \citep{Midlarsky1980, Neumayer2010, Mohler2013}, as in a ``diffusion'' processes \citep{Cohen1999, Forsberg2014}. Moreover, the activity of terrorism may also vary over time and often exhibits temporally clustered patterns like crime or insurgencies \citep{Anselin2000, Eck2005,Zammit2012}. 

Despite successful applications of space-time Bayesian models in similar fields of research, such as crime and conflict \citep{Zammit2012, Zammit2013, Mohler2013, Lewis2012, Rodrigues2010}, these models have not yet been applied in terrorism research. Most empirical research in terrorism has focused on its temporal dimension \citep{Hamilton1983, Porter2012, Brandt2012, Barros2003, Enders1999, Suleman2012, Bilal2012, Holden1986, Enders1993, Enders2011, Enders2005b, Enders2000, Weimann1988, Raghavan2013}, or has considered purely spatial models only \citep{Braithwaite2007, Savitch2001, Brown2004}. Moreover, studies that have explicitly integrated both space and time dimensions have been carried out at country or higher level of analysis \citep{LaFree2010, Midlarsky1980, Neumayer2010, Enders2006, Gao2013}, or at subnational level of analysis but within specific study areas \citep{LaFree2012, Behlendorf2012, Nunn2007, Piegorsch2007, Oecal2010, Medina2011, Siebeneck2009, Mohler2013}. 

As a result, scholars have failed to systematically capture the fine-scale spatial dynamics of terrorism. Local drivers of terrorism have not been identified and their effects have not been systematically assessed. In an effort to address this shortcoming, we use space-time Bayesian models based on the stochastic partial differential equation (SPDE) approach implemented through computationally efficient integrated nested Laplace approximation (INLA) techniques \citep{Rue2009, Lindgren2011}. Our approach, which integrates spatially explicit covariates and data on terrorist events, allows to capture local-scale spatial patterns of both the capacity of terrorist attack of causing death and the number of lethal attacks, which will respectively further refer to the \textit{lethality} of terrorism and the \textit{frequency} of lethal attacks.

Moreover, this study provides a measure of the effects of local-scale factors involved in the spatial and temporal variations of the lethality and the frequency of terrorism across the world from 2002 to 2013. The results of this study could benefit policy makers needing a systematic and accurate assessment of the security threat posed by deadly terrorist activity at subnational level. The paper is structured as follows. Section~\ref{sec:data} briefly introduces the data used for the analysis. The statistical models are described in Section~\ref{sec:model} and the results are provided in Section~\ref{sec:result}. Finally, conclusions and recommendations for further research are discussed in Section~\ref{sec:discussion}. The computer code used in this paper is available in supplementary material. 

\section{Data Selection}
\label{sec:data}
\subsection{Terrorism database}
\label{subsec:terrorismdatabase}
In order to build valuable, empirically-based models, it is crucial to base an analysis on a data source that is as suitable as possible for a given study \citep{Zammit2012}. There are currently four major databases that provide data on worldwide \textit{non-state} terrorism (terrorism perpetrated by non-state actors): the \textit{Global Terrorism Database} (GTD), the \textit{RAND Database of Worldwide Terrorism Incidents} (RDWTI), the \textit{International Terrorism: Attributes of Terrorist Events} (ITERATE), and the \textit{Global Database of Events, Language, and Tone} (GDELT). ITERATE has been extensively referred to in terrorism research \citep{Enders2011}, however, events are geolocalised at the country level, which does not allow to capture subnational processes. GDELT is not suitable for our purpose since it does not provide the number of fatalities or information on lethality of terrorism. Equally problematic, GDELT uses a fully automated coding system based on Conflict and Mediation Event Observations (CAMEO) (for further information on CAMEO, see: \url{http://data.gdeltproject.org/documentation/CAMEO.Manual.1.1b3.pdf}), which may lead to a strong geographic bias, as mentioned by \citet{Hammond2014}. 

Hence, RDWTI and GTD are the only potentially relevant databases that provide geolocalised terrorist events across the world. Drawing from \citeauthor{Sheehan2012}'s approach to compare terrorism databases (\citeyear{Sheehan2012}), we defined four criteria to select the one which will be used in our study: \textit{conceptual clarity}, \textit{scope}, \textit{coding method}, and \textit{spatial accuracy}. Given that the concept of terrorism is intrinsically ambiguous and being debated to this day \citep{Beck2013}, \textit{conceptual clarity} in both the definition of terrorist events and the \textit{coding method} used to gather data are crucial. In both GTD and RDWTI, the definition used to class an event as terrorism is clearly specified. The \textit{coding method} of GTD appears more rigorous, since events are gathered from numerous sources and articles (more than 4,000,000 news articles and 25,000 news sources used from 1998 and 2013 \citep{GTD2014}), which limits the risk of bias resulting from inconsistent ways of reporting the number of fatalities in different media \citep{Drakos2006a, Drakos2007}. The data collection methodology used in RDWTI is less reliable since some events are gathered from two sources only. Moreover, the \textit{scope} of GTD is wider than RDWTI. GTD is updated annually and includes more than 140,000 events from 1970 until 2014, whereas RDWTI was not updated after 2009 and includes 40,129 events from 1969 to 2009 only \citep{Start2014, Rand2011}.

Since this research investigates fine-scale spatial phenomena, we put particular emphasis on the \textit{spatial accuracy} of the data. GTD is the only database that includes a variable assigning the spatial accuracy of each individual observation. Spatial accuracy is represented by an ordinal variable called \textit{specificity}, with 5 possible levels of spatial accuracy (for further information on \textit{specificity}, see GTD codebook: \url{https://www.start.umd.edu/gtd/downloads/Codebook.pdf})
\citep{GTD2014}. Based on all these considerations we have chosen GTD as the appropriate data source for this study
. The dataset contains 35,917 spatially accurate events (events corresponding to the highest level of spatial accuracy, with \textit{specificity}=1), occurring between 2002 and 2013. 

\subsection{Covariates}
\label{subsec:covariates}

Potentially relevant covariates were identified based on a thorough review of 43 studies carried out at country level by \citet{Gassebner2011}, which highlighted the main explanatory factors among 65 potential determinants of terrorism. Among those factors, we consider covariates that satisfy two essential characteristics: (i) potential relationship with the lethality of terrorism and/or the frequency of lethal terrorist attacks; (ii) 
availability at high spatial resolution, in order to model fine-scale spatial dynamics of terrorism worldwide. Seven spatial and space-time covariates met these criteria and their potential association with terrorism is described in more detail below: satellite night light ($lum$), population density ($pop$), political regime ($pol$), altitude ($alt$), slope ($slo$), travel time to the nearest large city ($tt$), and distance to the nearest national border ($distb$). 

First, we assess the role of economic factors on the lethality of terrorism, whose possible effects are still under debate. Most country-level empirical studies have not provided any evidence of a linear relationship between terrorism and gross domestic product (GDP) \citep{Abadie2006, Drakos2006a, Gassebner2011, Krueger2008, Krueger2003, Piazza2006}, without excluding possible non-linear relationship \citep{Enders2012}. Case studies focused in the Middle East, including Israel and Palestine, showed that GDP is not significantly related to the number of suicide terrorist attacks \citep{Berman2008}. Few studies, however, found that countries with high per capita GDP may encounter high levels of terrorist attacks \citep{Tavares2004, Blomberg2009}
. In line with the subnational nature of our study, we use NOAA satellite lights at night (Version 4 DMSP-OLS) as a covariate, which provides information about worldwide human activities on a yearly basis and at a high spatial resolution (30 arc-second grid) \citep{Chen2011, NOAA2014}. This variable has been used as a proxy for socio-economic development measures such as per capita GDP estimation \citep{Sutton2002, Sutton2007, Elvidge2007, Ebener2005, Henderson2009}. Note that three versions are available: \textit{Average Visible}, \textit{Stable Lights}, and \textit{Cloud Free Coverages}. We use \textit{Stable Lights}, which filter background noise and identify zero-cloud free observations \citep{NOAA2014}. In order to compare values of different years, we calibrate the data according to the procedure described in \citet[Chap.6]{Elvidge2013}.

Second, we assess the role of demography. Cities may provide more human mobility, anonymity, audiences and a larger recruitment pool in comparison to rural areas (\citealp[p.~115]{Crenshaw1990}; \citealp{Savitch2001}). Large cities, in particular, offer a high degree of anonymity for terrorists to operate \citep[p.~41]{Laqueur1999}. More specifically, densely populated areas appear vulnerable and are usually more prone to terrorism than sparsely populated areas \citep{Ross1993, Savitch2001, Crenshaw1981, Swanstrom2002, Coaffee2010}. In addition, locations that shelter high-value symbolic targets (buildings or installations), human targets (government officials, mayors, etc.), and public targets (public transports, shopping centres, cinemas, sport arenas, public venues, etc.) are particularly vulnerable to suicide terrorism \citep[p.~167]{Hoffman2006}. Therefore, we use the Gridded Population of the World (v3), which provides population density on a yearly basis and at high-resolution (2.5 arc-minute grid) \citep{CIESIN2005}. Moreover, terrorists usually require free and rapid movement by rail or road in order to move from and to target points (\citealp{Heyman1980}; \citealp[p.~189]{Wilkinson1979}). We compute the travel time from each terrorist event to the nearest large city (more than 50,000 inhabitants) based on Travel Time to Major Cities \citep{Nelson2008} at a high spatial resolution (30 arc-second grid). 

Third, we assess the role of geographical variables: altitude, surface topography (slope), and distance to the nearest national border. Although the relationship between altitude, slope, and the lethality of terrorism is not straightforward, both variables provide an indication of the type of the geographical location, which could be a determining factor for terrorists regarding their choice of target \citep{Ross1993}. Moreover, \citet{Nemeth2014} suggested that distance to the nearest national border and altitude or slope might have an impact on terrorist activity. We extract both variables from NOAA Global Relief Model (ETOPO1), which provides altitude values at high spatial resolution (1 arc-minute grid) \citep{Amante2009}.

Fourth, we assess the role of democracy. Under-reporting biases may occur especially in non-democratic countries where the press is often not free \citep{Drakos2006a, Drakos2007}. We extract the level of democracy from Polity IV Project, Political Regime Characteristics and Transitions, 1800-2014 (\text{Polity IV}) \citep{Marshall2014}. \text{Polity IV} informs about the level of freedom of press, and captures the level of democracy from $-10$ (hereditary monarchy) to $+10$ (consolidated democracy) for most independent countries from 1800 to 2014. Therefore, it has been commonly referred as proxy for measuring the type of regime or the extent of constraints in democratic institutions \citep{Gleditsch2007,Li2005,Piazza2006}.

Fifth, we assess the role of ethnicity. We compute the number of different ethnic groups from the ``Georeferencing of ethnic groups'' (GREG) database. GREG is the digitalised version of the Soviet Atlas Narodov Mira (ANM), and counts 1,276 ethnic groups around the world \citep{Weidmann2010b}. Although ANM includes information dating back to the 1960s, it is still regarded as a reliable source for ethnicity across the world \citep{Bhavnani2012, Morelli2014}. Although ethnic diversity does not necessarily lead to violence per se \citep[p.~68]{Silberfein2003}, studies at country-level suggest that more terrorism may occur in ethnically fragmented societies \citep{Kurrild2006, Gassebner2011}, in countries with strong ethnic tensions, or may originate from ethnic conflicts in other regions \citep{Basuchoudhary2010}. 

\section{Modelling the Spatial Dynamics of Terrorism}
\label{sec:model}
\subsection{SPDE Framework}
\label{subsec:spdemodel}
\begin{figure}[hb]
\centering
\raisebox{.2\height}{\includegraphics[scale=0.17]{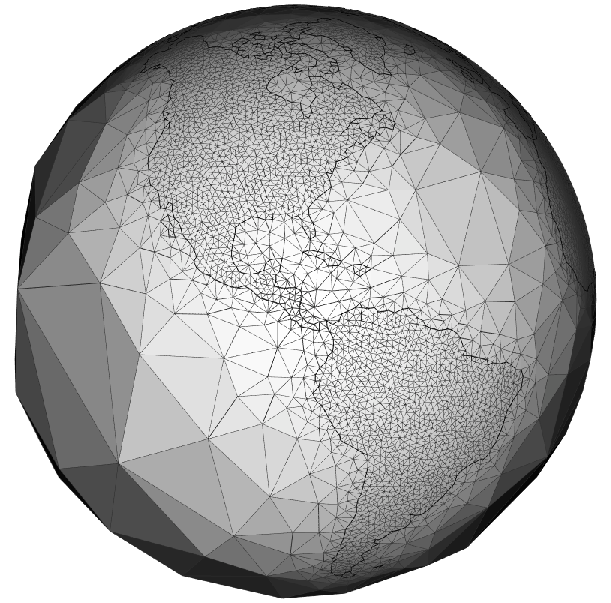}} 
\includegraphics[scale=0.17]{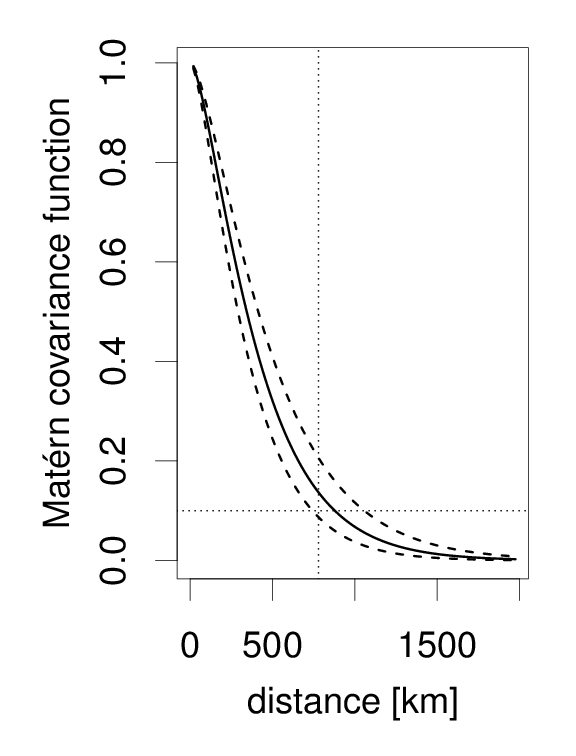}
\includegraphics[scale=0.17]{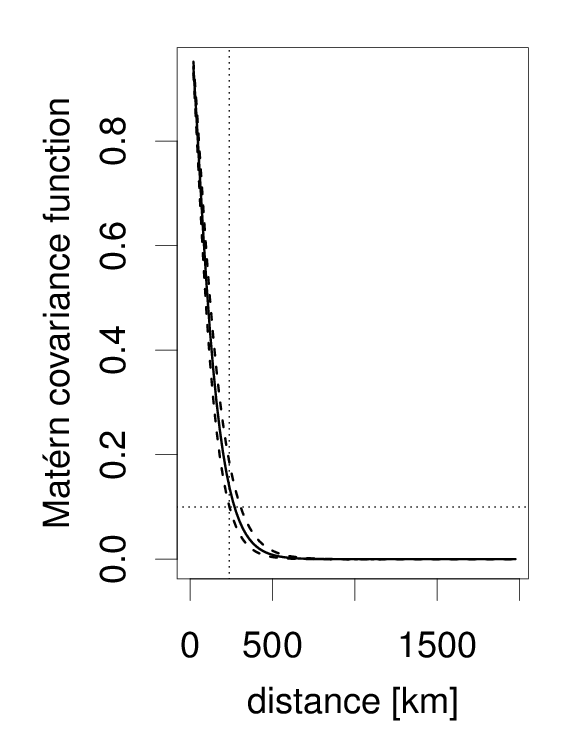}
\caption{\textit{Left}: Constrained Refined Delaunay Triangulation (mesh) with 9,697 vertices, on the top of which the SPDE and its discretised solutions (GMRF) are built \citep{Lindgren2012}. Mat\'{e}rn correlation function (solid line) with parameters $\lambda=1$ and its $CI_{95\%}$ (dashed lines) for Bernoulli model with posterior mean $\kappa\approx 23.32$ (\textit{centre}) and Poisson model with posterior mean $\kappa\approx 76.03$ (\textit{right}). Note the differences in the posterior mean range (vertical dotted line) between Bernoulli (\textit{centre}) and Poisson models (\textit{right}), which corresponds to correlation $\approx 0.1$ (horizontal dotted line)} %
\label{fig:mesh}
\end{figure}

We assume that both the lethality of terrorism and the frequency of lethal terrorist attacks are continuous phenomena with Gaussian properties, which exhibit dependencies in both space and time. We suggest modelling their spatial dynamics through the SPDE approach introduced by \citet{Lindgren2011}. The solution of the SPDE given in equation~(\ref{eq:spde}) is a Gaussian field (GF) \citep{Lindgren2011}, whose approximation represents a Gaussian Markov random field (GMRF) used herein to model the spatio-temporal dependencies inherent in the data. The linear SPDE can be formulated as:
\begin{equation}
(\kappa^2 - \Delta )^{\alpha/2}(\tau \zeta(\bm{s})) = \mathcal{W}(\bm{s}),\quad \bm{s}\in \mathcal{D}\;,\label{eq:spde}
\end{equation}
with the Laplacian $\Delta$, smoothness parameter $\alpha=\lambda+1$ (for two-dimensional processes), scale parameter $\kappa>0$, variance parameter $\tau$, domain $\mathcal{D}$, and Gaussian spatial white noise $\mathcal{W}(\bm{s})$ \citep[chap 6]{Blangiardo2015}. The stationary solution of equation~(\ref{eq:spde}) is the GF ($\zeta(\bm{s}$)) with Mat\'{e}rn covariance function:
\begin{equation}
Cov(\zeta(\bm{s}_{\bm{i}}),\zeta(\bm{s}_{\bm{j}})) =
\sigma^2_{\zeta}\frac{1}{\Gamma(\lambda)2^{\lambda-1}}\bigg(\kappa\left\|\bm{s}_{\bm{i}}-\bm{s}_{\bm{j}}\right\|\bigg)^\lambda K_\lambda\bigg(\kappa\left\|\bm{s}_{\bm{i}}-\bm{s}_{\bm{j}}\right\|\bigg)
\;,\label{eq:cov}
\end{equation}

where $\left\|\bm{s}_{\bm{i}}-\bm{s}_{\bm{j}}\right\|$ is the Euclidean distance between two locations, $\sigma^2_{\zeta}$ is the marginal variance, and $K_\lambda$ is the modified Bessel function of the second kind and order $\lambda>0$. The distance from which the spatial correlation becomes negligible (for $\lambda>0.5$) is given by the range $r$ (vertical dotted line in figure~\ref{fig:mesh}, \textit{centre} and \textit{right}), which can be empirically derived from the scale parameter $r=\sqrt{8\lambda}/\kappa$ to be estimated. The GF ($\zeta(\bm{s}$)) is approximated as a GMRF ($\tilde{\zeta}(\bm{s})$) through a finite element method using basis functions defined on a Constrained Refined Delaunay Triangulation (mesh) over the earth, modelled as a sphere (figure~\ref{fig:mesh}, \textit{left}) \citep{Lindgren2011}. Here, we use a three-stage Bayesian hierarchical modelling framework \citep{Banerjee2014} to model the lethality of terrorism as a Bernoulli process (Section~\ref{subsec:bernoullimodel}) and the frequency of lethal terrorist attacks as a Poisson process (Section~\ref{subsec:poissonmodel}).
\vspace{0pt}

\subsection{Bernoulli Model}
\label{subsec:bernoullimodel}
While terrorist attacks are inherently discrete in space (terrorist attacks occur at specific locations on earth), we consider their lethality $Y$ as a continuous phenomenon (in the sense of geostatistics \citep{Cressie1991}), which is observed at $\bm{s}$ locations (attacks) over the surface of the earth $\mathbb{S}^2$ at time $t\in \mathbb{R}$. The lethality is assumed to be the realisations of a continuously indexed space-time process $Y(\bm{s},t)\equiv \{y(\bm{s},t): (\bm{s}, t) \in \mathcal{D} \subseteq \mathbb{S}^2 \times \mathbb{R}\}$, from which inference can be made about the process at any desired locations in $\mathcal{D}$ \citep{Cameletti2013}. Hence, our hierarchical modelling framework is composed of three levels:
\begin{subequations}
\begin{align}
y(\bm{s}_i,t)\vert \bm{\theta}, \tilde{\zeta}\sim \text{Bernoulli($\pi(\bm{s}_i,t)$)}\label{eq:BHM1} \\
\text{logit($\pi(\bm{s}_i,t)$)}\vert \bm{\theta}=\eta(\bm{s}_i,t)\vert \bm{\theta}=\beta_{0} + \bm{z}(\bm{s}_i,t)\bm{\beta} + \tilde{\zeta}(\bm{s}_i,t) + \epsilon(\bm{s}_i,t) \label{eq:BHM2} \\
\bm{\theta}=p(\bm{\theta}),\label{eq:BHM3}
\end{align}
\label{eq:BHM}
\end{subequations}

where the lethality $y(\bm{s}_i,t)$ is a dichotomous variable that takes the value 1 if the attack generated one or more deaths, and 0 if not (equation~(\ref{eq:BHM1})). The parameters to be estimated are $\bm{\theta}=\{\beta_{0}, \bm{\beta}, \sigma^2_\epsilon,\tau,\kappa,\rho\}$, which include the precision of the GMRF $\tau=1/\sigma^2_{\tilde{\zeta}}$, the scale parameter $\kappa>0$ of its Mat\'{e}rn covariance function (equation~(\ref{eq:cov})), and the temporal autocorrelation parameter $\left|\rho\right|<1$ (described in more detail below) (equations~(\ref{eq:BHM1}),~(\ref{eq:BHM2}),~(\ref{eq:BHM3})). The conditional distribution of the linear predictor $\eta(\bm{s}_i,t)=\;$logit($\pi(\bm{s}_i,t)$), given the parameters $\bm{\theta}$ (equation~(\ref{eq:BHM2})), includes Gaussian white noise $\epsilon(\bm{s}_i,t)\sim \mathcal{N}(0,\sigma^2_\epsilon)$, with measurement error variance $\sigma^2_\epsilon$, a vector of $k$ covariates $\bm{z}(\bm{s}_i,t)=(z_1(\bm{s}_i,t),\dots, z_k(\bm{s}_i,t))$ with coefficient vector $\bm{\beta}=(\beta_1,\dots,\beta_k)^{\prime}$ and the GMRF $\tilde{\zeta}(\bm{s}_i,t)$.

Based on GTD \citep{GTD2014}, we extract the lethality of 35,917 accurately geolocalised terrorist attacks that occurred from year $t=2002$ to $t=2013$ in $i=1,\dots,35,917$ space-time locations $\bm{s}_i$. In equation~(\ref{eq:BHM1}), we assume that $y(\bm{s}_i,t)$ follows a conditional Bernoulli distribution with probability $\pi(\bm{s}_i,t)$ of observing a lethal event and $1-\pi(\bm{s}_i,t)$ of observing a non-lethal event, given the GMRF $\tilde{\zeta}$ and parameters $\bm{\theta}$.

In order to minimise the complexity of the models, and consequently, reduce the computing time required to fit them, we assume a separable space-time covariance \citep[chap 7]{Blangiardo2015}. Hence, the GMRF $\tilde{\zeta}(\bm{s}_i,t)$ follows a first-order autoregressive process AR(1): $\tilde{\zeta}(\bm{s}_i,t)=\rho\;\tilde{\zeta}(\bm{s}_i,t-1)+ \psi(\bm{s}_i,t)$, time independent zero-mean Gaussian field $\psi(\bm{s}_i,t)$ with $Cov(\tilde{\zeta}(\bm{s}_i,t),\tilde{\zeta}(\bm{s}_j,u)) = 0$ if $t\neq u$, and $Cov(\tilde{\zeta}(\bm{s}_i),\tilde{\zeta}(\bm{s}_j))$ if $t=u, \forall t, u\in \{2002,\dots,2013\}$. 

Prior distributions (equation~(\ref{eq:BHM3})) are set for the parameters to be estimated ($\bm{\theta}=\{\beta_{0}, \bm{\beta}, \sigma^2_\epsilon,\tau,\kappa,\rho\}$), which includes GMRF's precision $\tau=1/\sigma^2_{\tilde{\zeta}}$, scale parameter $\kappa>0$ of its Mat\'{e}rn covariance function (equation~(\ref{eq:cov})), and temporal autocorrelation parameter $\left|\rho\right|<1$. We use the default option for the stationary model in \texttt{R-INLA} as priors on $\tau$ and $\kappa$ through prior distribution on $\log(\tau), \log(\kappa) \sim\mathcal{N}(0,1)$, so that $\log(\kappa(\bm{s})) =\log(\kappa)$ and $\log(\tau(\bm{s}))=\log(\tau)$. For prior sensitivity analysis, we compare our results with alternative priors on $\tau$ and $\kappa$, which is further discussed in Section~\ref{sec:result}. 

The modelling approach is used to identify regions of abnormally high values (hot-spots) of the lethality of terrorism from the estimated posterior distribution of the probability of lethal attack interpolated in all space-time locations $\bm{s},t \in \mathcal{D}$. For each year ($t=2002,\dots,2013$), we identify locations where the 95\% credible interval (CI) for the probability of lethal attack is above a threshold $\epsilon$ ($0<\epsilon<1$): 
\begin{equation}
\label{eq:hotspot}
L_{CI_{95\%}} \pi(\bm{s},t)>\epsilon, \quad \bm{s},t \in \mathcal{D}, 
\end{equation}
where $L_{CI_{95\%}}$ is the lower bound of the 95\% CI, $\pi(\bm{s},t)$ is the probability of the attack(s) to be lethal in location $s$ and time $t$, $\epsilon$ is the threshold, and $\mathcal{D}$ is the domain. We define a hot-spot as the groups of contiguous locations $\bm{s}$ that satisfy equation~(\ref{eq:hotspot}). Here we use $\epsilon=0.5$, which means that we define hot-spot areas where it is more likely to have lethal attacks than non-lethal attacks. In other terms, we are 95\% confident that the true value of the probability of lethal attack is greater than 50\%. 

\subsection{Poisson Model}
\label{subsec:poissonmodel}
In addition to modelling the lethality of terrorist attacks, one might identify locations that are more likely to encounter a higher number of lethal attacks over a year, which we further refer as the \textit{frequency} of lethal terrorist attacks. The identification of such locations could be crucial for city planners, emergency managers, insurance companies, and property administrators for example, since this information could be used to better allocate resources used to prevent and counter terrorism \citep{Piegorsch2007,Nunn2007}. As in equation~(\ref{eq:BHM}), we use a three-stage Bayesian hierarchical modelling framework \citep{Banerjee2014}:
\begin{subequations}
\begin{align}
y(\bm{s}_i,t)\vert \bm{\theta}, \tilde{\zeta}\sim \text{Poisson($\mu(\bm{s}_i,t)$)} \label{eq:PHM1} \\
\text{log($\mu(\bm{s}_i,t)$)}\vert \bm{\theta}=\eta(\bm{s}_i,t)\vert \bm{\theta}=\beta_{0} + \bm{z}(\bm{s}_i,t)\bm{\beta} + \tilde{\zeta}(\bm{s}_i,t) + \epsilon(\bm{s}_i,t)\label{eq:PHM2} \\
\bm{\theta}=p(\bm{\theta}). \label{eq:PHM3}
\end{align}
\label{eq:PHM}
\end{subequations}

Based on \citet{GTD2014}, we consider the observed number of lethal attacks ($y(\bm{s}_i,t)$ in equation~(\ref{eq:PHM1})) that occurred in a period of 12 years ($t=2002,\dots,2013$) in 6,386 locations $\bm{s}_i$ within a $0.5\degree$ radius of cities' centroids. Since we model a ``count'' variable ($y(\bm{s}_i,t)$), it is convenient to aggregate events that occurred in very close locations within identical municipality areas for example. This has resulted in spatial aggregation reducing the number of observations from 35,917 (equation~(\ref{eq:BHM})) to 6,386 (equation~(\ref{eq:PHM})). Moreover, we assume that $y(\bm{s}_i,t)$ follows a Poisson distribution with parameter $\mu(\bm{s}_i,t)$, with $\log(\mu(\bm{s}_i,t))=\eta(\bm{s}_i,t)$ and expected value $\mathbb{E}(y(\bm{s}_i,t))= \mu(\bm{s}_i,t)$.

Equation~(\ref{eq:PHM2}) is structurally identical to equation~(\ref{eq:BHM2}). As in the Bernoulli model (Section~\ref{subsec:bernoullimodel}), we use the default option for the stationary model in \texttt{R-INLA} as priors on $\tau$ and $\kappa$ through prior distribution on $\log(\tau), \log(\kappa) \sim\mathcal{N}(0,1)$, so that $\log(\kappa(\bm{s})) =\log(\kappa)$ and $\log(\tau(\bm{s}))=\log(\tau)$. To assess prior sensitivity, we compare our results with elicited priors on $\kappa$ and $\tau$, which is further discussed in Section~\ref{sec:result}. As in the Bernoulli model, we identify hot-spots of high number of lethal attacks by replacing the posterior probability $\pi$ (equation~(\ref{eq:hotspot})) with the posterior expected number of lethal events $\mu(\bm{s}_i,t)$ (equation~(\ref{eq:PHM1})). We set $\epsilon=5$, which highlights cells with an expected number of lethal attacks greater than 5 (Figure~\ref{fig:hotspot}). This threshold corresponds to the 90\textsuperscript{th} percentile of the number of lethal attacks observed in the sample ($n=6,386$).

\section{Results}
\label{sec:result}

\subsection{Explaining the Spatial Dynamics of Terrorism}
\label{subsec:explain}
We use INLA as an accurate and computationally efficient model fitting method \citep{Rue2009, Held2010, Simpson2016}
. With a 12-core Linux machine (99 GB of RAM), \texttt{R-INLA} requires approximately 7 days to fit each model. It is likely that fitting a similar model with MCMC methods would take too long to be practically feasible. We select one Bernoulli (among models with 0, 1, 2, 3, 4, and 5 covariates) and one Poisson model (among models with 3, 4, and 5 covariates), which exhibit the lowest Deviance Information Criteria (DIC) and Watanabe-Akaike information criterion (WAIC) \citep{Watanabe2010, Spiegelhalter2002}. The selected Bernoulli model includes $k=4$ standardised covariates with corresponding coefficients: satellite night light ($\beta_{lum}$), altitude ($\beta_{alt}$), ethnicity ($\beta_{greg}$), and travel time to the nearest large city ($\beta_{tt}$). The selected Poisson model includes $k=5$ standardised covariates with corresponding coefficients: satellite night light ($\beta_{lum}$), altitude ($\beta_{alt}$), democracy level ($\beta_{pol}$), population density ($\beta_{pop}$), and travel time to the nearest large city ($\beta_{tt}$). 

The Bernoulli model (table~\ref{tab:bincoef}, columns ``Bernoulli'') suggests that terrorist attacks are more likely to be lethal far away from large cities, in higher altitude, in locations with higher ethnic diversity ($CI_{95\%}\;\beta_{tt}, \beta_{greg}, \beta_{alt}>0$), and are less likely in areas with higher human activity ($CI_{95\%}\;\beta_{lum}<0$). As an illustration, we compare the effect of an hypothetical 50\% increase in the mean of luminosity ($lum$) on the probability of encountering lethal attacks ($\pi$). Since the covariates $\bm{\beta}$ are standardised, particular care should be taken to estimate such effect. Recall that NOAA satellite data on lights at night provide non-standardised values of luminosity from 0 (min) to 63 (max) \citep{NOAA2014}. Based on 35,917 observations used in the Bernoulli model, a 50\% increase in the mean of luminosity corresponds to an increase from 34.3 to 51.5, or equivalently, from 0 to 0.73 on a standardised scale. In the benchmark scenario, all predictors, including $lum$ and other covariates $tt$, $greg$, $alt$, and the GMRF ($\tilde{\xi}$) are held equal to 0. Hence, the linear predictor $\eta$ (equation~(\ref{eq:BHM2})) equals 0 and $\pi=1/(1+\exp{(-\eta)})=1/(1+1)=0.5$. In the scenario including a 50\% increase in the mean of luminosity, $lum=0.73$, $\eta=\beta_{lum}\times lum=-0.11\times0.73=-0.0803$, given that $\beta_{lum}=-0.11$ (table~\ref{tab:bincoef}, columns ``Bernoulli''). Hence, $\pi=1/(1+\exp{-(-0.0803)})\cong0.48$. As a result, an increase in the mean of luminosity by 50\% decreases the probability of lethal attacks by approximately 2\%.

In contrast, the Poisson model (table~\ref{tab:bincoef}, columns ``Poisson'') suggests that more economically developed areas, and locations with higher democratic levels ($CI_{95\%}\;\beta_{lum}, \beta_{pol}>0$) and close to large cities ($CI_{95\%}\;\beta_{tt}<0$) are more likely to encounter a higher number of lethal attacks. However, we did not find a significant relationship between altitude, population density, and the number of lethal attacks ($0 \in CI_{95\%}\; \beta_{alt}, \beta_{pop}$). The interpretation of these results is discussed in further detail in Section~\ref{sec:discussion}. Note that the results from the two models cannot be directly compared since they are based on a different number of observations, set of covariates, and spatial aggregation.

As with the Bernoulli model, we analyse the effect of a 50\% increase in the mean of luminosity on the expected number of lethal attacks ($\mu$) estimated by the Poisson model, with all predictors, including $lum$ and the other covariates $tt$, $alt$, $pol$, $pop$, and the GMRF ($\tilde{\xi}$) held equal to 0. In the benchmark scenario, $\eta$ (equation~(\ref{eq:PHM2})) equals to 0, therefore the expected number of lethal attacks $\mu=\exp{(0)}=1$. Based on 6,386 observations used in the Poisson model, a 50\% increase in the mean of luminosity corresponds to an increase from 7.45 to 11.2, or equivalently, from 0 to 0.25 in the corresponding standardised scale. Hence, $\eta=(\beta_{lum}\times lum)$, with $\beta_{lum}=0.51$ (table~\ref{tab:bincoef}, columns ``Poisson''). We obtain $\mu=\exp{(0.51\times 0.25)}\cong1.14$. Therefore, an increase in the mean of luminosity by 50\% increases the expected number of lethal attacks by approximately 0.14.
\begin{table}[h!]
\caption{\label{tab:bincoef}Posterior mean, standard deviation, and 95\% credible intervals (CI) of the intercept $\beta_0$, the coefficients of the standardised covariates ($\bm{\beta}$), the temporal ($\rho$), and the spatial parameters of the GMRF $\tilde{\zeta}$ ($\kappa$, $\sigma^2_{\tilde{\zeta}}$, and range $r$) estimated in the Bernoulli ($n=35,917$) and Poisson ($n=6,386$) models. Note that the 95\% CI associated with the spatial parameters correspond to 95\% highest probability density intervals.}
\centering
\fbox{%
\begin{tabular}{lcccccc}
& \multicolumn{3}{l}{Bernoulli (n=35,917)}&  \multicolumn{3}{l}{Poisson (n=6,386)} \\
\cmidrule(l){2-4}  \cmidrule(l){5-7}
&  mean&sd& 95\% CI  &mean&sd& 95\% CI \\
Cov. ($\beta$) & &  &  &&& \\
\cmidrule(l){1-1}
$\beta_0 $    & -0.58 & 0.13 & (-0.83; -0.32)      & -1.37 & 0.09 & (-1.55; -1.19)\\
$\beta_{lum}$&-0.11& 0.02  &    (-0.15; -0.06) &0.51& 0.02  &    (\;0.47; \;0.54)\\
$\beta_{tt}$&0.06& 0.02    & (\;0.02; \;0.09)         &-0.38& 0.02    & (-0.42; -0.34)\\
$\beta_{greg}$ &0.04 &0.02 &  (0.003; 0.08)      &  &  & \\
$\beta_{alt}$ & 0.08& 0.03&  (\;0.03; \;0.13)          & 0.03& 0.02&  (-0.001; 0.07)\\
$\beta_{pol}$   &  &  & & 0.42 & 0.01 &(\;0.39; \;0.45)\\
$\beta_{pop}$   &  &  & & 0.009 & 0.02 &(-0.03; \;0.04)\\
\\
GMRF ($\tilde{\zeta}$)  & &   &  &&&\\
\cmidrule(l){1-1}
$\rho$ (AR1) & 0.91 & 0.01 & (0.88; 0.91)  & 0.91 & 0.07 & (0.90; 0.93)\\
$\kappa$ & 23.3 & & (19.4; 27.6) & 76.0 & & (66.9; 85.8) \\
$\sigma^2_{\tilde{\zeta}}$ & 2.27 & & (1.83; 2.73)  & 4.61 & & (4.11; 5.13) \\
$r$ $\left[km\right]$ & 779 &  & (643; 915)  & 238 &  & (208; 267) \\
\\
\end{tabular}}
\end{table}

\subsection{Quantifying the Uncertainty}
\label{subsec:uncertainty}

\begin{sidewaysfigure}
\centering
\begin{subfigure}{0.32\textheight}
\includegraphics[width=0.32\textheight,height=6cm]{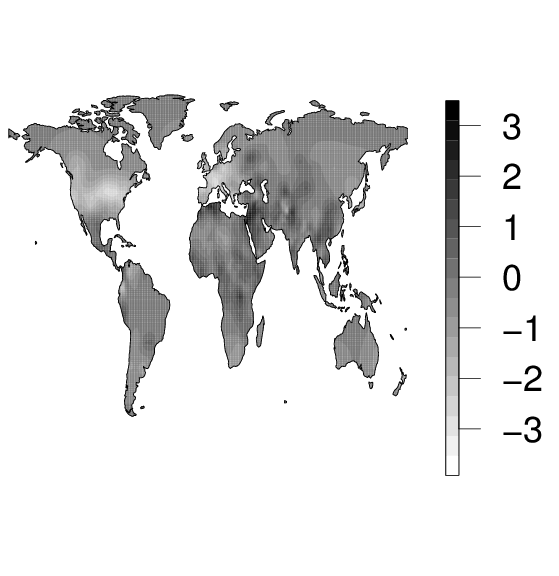}
\vspace{-2em}
\caption{posterior mean $\tilde{\zeta}(\bm{s},2002)$}
  \label{fig:Bbirf1}
\end{subfigure}
\vspace{0.8em}
\begin{subfigure}{0.32\textheight}
\includegraphics[width=0.32\textheight,height=6cm]{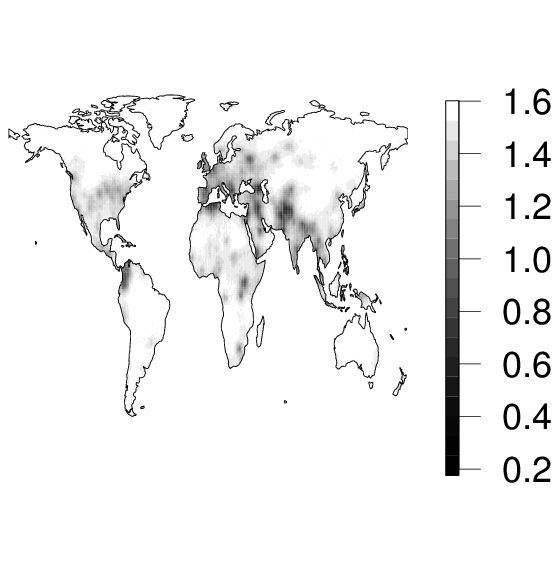} 
\vspace{-3em}
\caption{posterior standard deviation $\sigma_{\tilde{\zeta}(\bm{s},2002)}$}
  \label{fig:Bbisdrf1}
\end{subfigure}
\vspace{0.5em}
\begin{subfigure}{0.32\textheight}
\includegraphics[width=0.32\textheight,height=6cm]{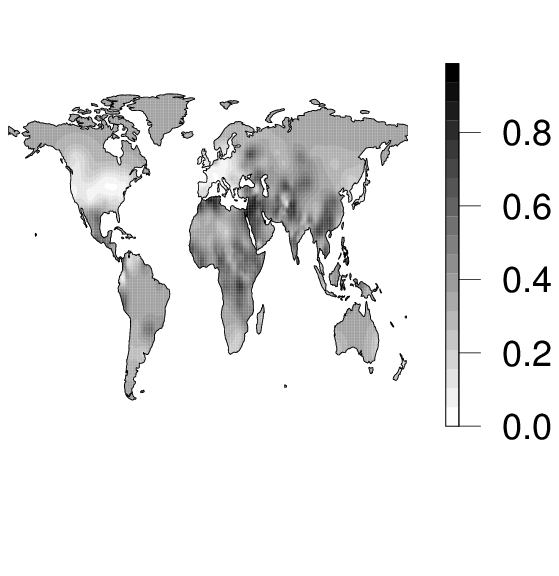} 
\vspace{-3.2em}
\caption{posterior mean $\pi(\bm{s},2002)$}
  \label{fig:Bprobsurf1}
\end{subfigure}

\vspace{-2.5em}
\begin{subfigure}{0.32\textheight}
\includegraphics[width=0.32\textheight,height=6cm]{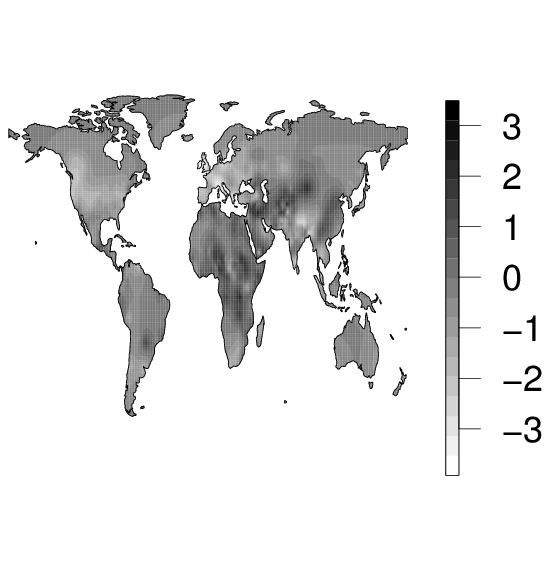}
\vspace{-2em}
\caption{posterior mean $\tilde{\zeta}(\bm{s},2013)$}
  \label{fig:Bbirf12}
\end{subfigure}
\vspace{0.3em}
\begin{subfigure}{0.32\textheight}
\includegraphics[width=0.32\textheight,height=6cm]{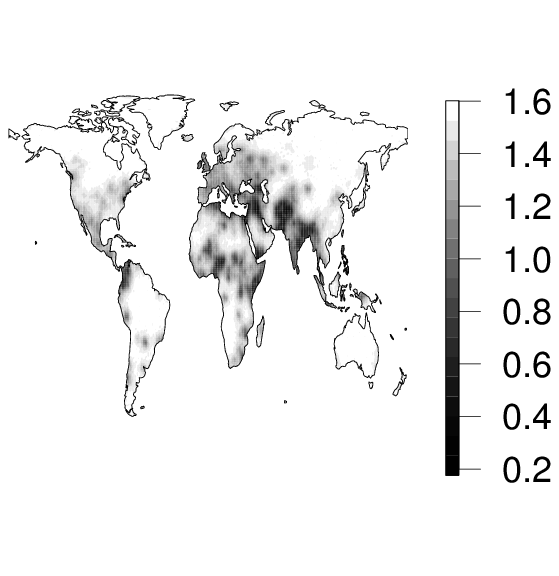} 
\vspace{-2.9em}
\caption{posterior standard deviation $\sigma_{\tilde{\zeta}(\bm{s},2013)}$}
  \label{fig:Bbisdrf12}
\end{subfigure}
\vspace{0em}
\begin{subfigure}{0.32\textheight}
\includegraphics[width=0.32\textheight,height=6cm]{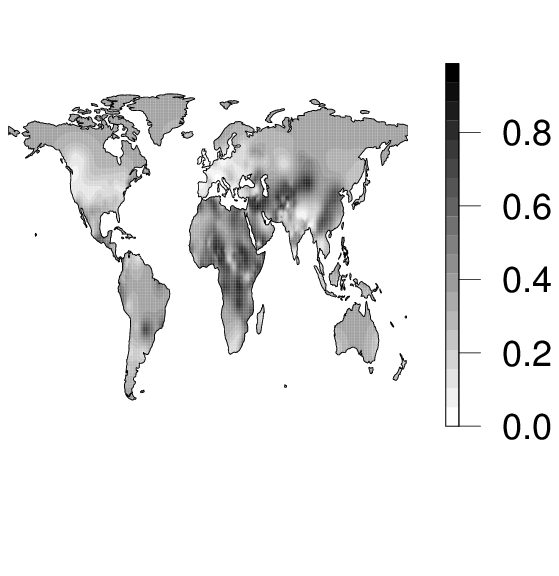} 
\vspace{-3.2em}
\caption{posterior mean $\pi(\bm{s},2013)$}
  \label{fig:Bprobsurf12}
\end{subfigure}
\vspace{1em}
\caption{Bernoulli model of the lethality of terrorist attacks with GMRF posterior mean $\tilde{\zeta}(\bm{s},t)$ (\textit{left}), posterior standard deviation $\sigma_{\tilde{\zeta}(\bm{s},t)}$ (\textit{centre}), and posterior mean probability of lethal attack $\pi(\bm{s},t)$ (\textit{right}) estimated in the 9,697 locations of mesh vertices and interpolated in all locations on land surface $\bm{s} \in \mathbb{S}^2$. Illustrative projected maps provide values on land surface for years $t=2002$ (\textit{top}) and $t=2013$ (\textit{bottom}). Note the presence of high uncertainty expressed through high values of $\sigma_{\tilde{\zeta}_{s,t}}$ in e.g. Siberia or Amazonian areas due to the sparsity or absence of terrorist events (\textit{top centre and bottom centre}). Moreover, one can observe an increase in the probability of lethal attack ($\pi_{s,t}$) from 2002 (\textit{top right}) to 2013 (\textit{bottom right}) in some African and Middle East areas}
  \label{fig:prob}
\end{sidewaysfigure}

\begin{sidewaysfigure}
\centering

\begin{subfigure}{0.32\textheight}
\includegraphics[width=0.32\textheight,height=6cm]{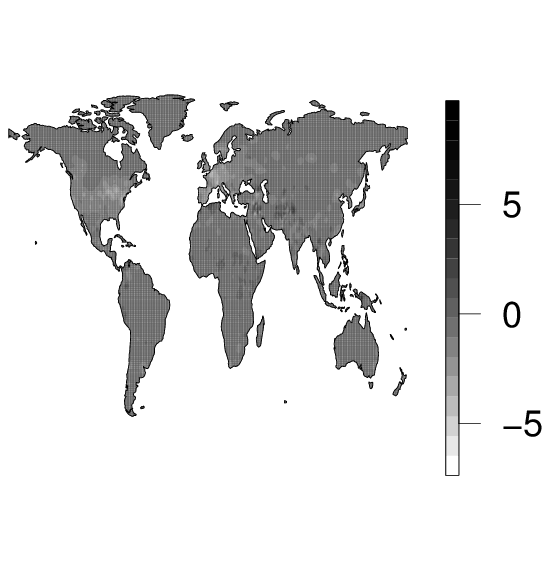}
\vspace{-2.2em}
\caption{posterior mean $\tilde{\zeta}(\bm{s},2002)$}
  \label{fig:Pbirf1}
\end{subfigure}
\vspace{1em}
\begin{subfigure}{0.32\textheight}
\vspace{0.2em}
\includegraphics[width=0.32\textheight,height=6cm]{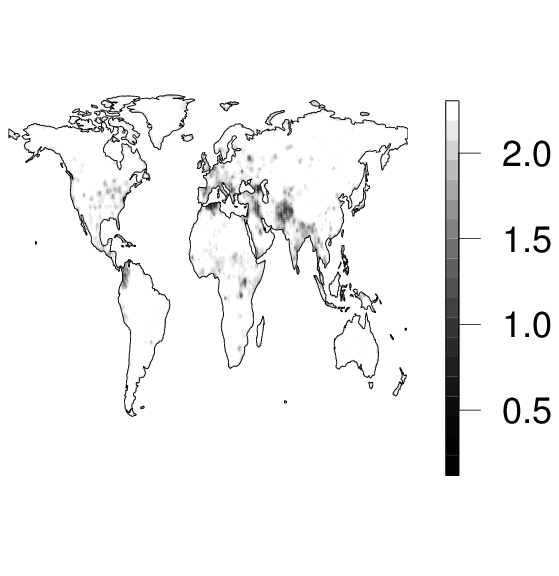} 
\vspace{-3.5em}
\caption{posterior standard deviation $\sigma_{\tilde{\zeta}(\bm{s},2002)}$}
  \label{fig:Pbisdrf1}
\end{subfigure}
\begin{subfigure}{0.32\textheight}
\vspace{0.2em}
\includegraphics[width=0.32\textheight,height=6.2cm]{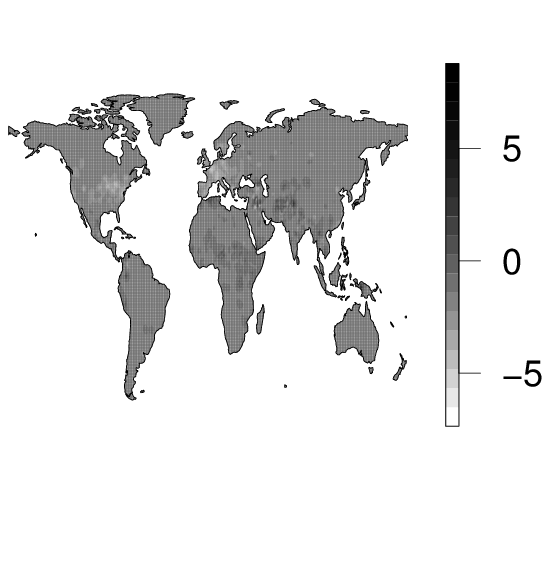} 
\vspace{-4.3em}
\caption{posterior mean log($\mu(\bm{s},2002)$)}
  \label{fig:Pprobsurf1}
\end{subfigure}

\vspace{-2.0em}
\begin{subfigure}{0.32\textheight}
\includegraphics[width=0.32\textheight,height=6cm]{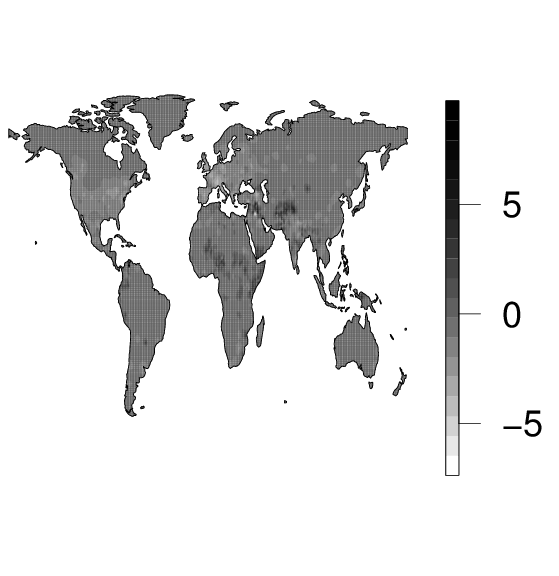}
\vspace{-2em}
\caption{posterior mean $\tilde{\zeta}(\bm{s},2013)$}
  \label{fig:Pbirf12}
\end{subfigure}
\vspace{1em}
\begin{subfigure}{0.32\textheight}
\includegraphics[width=0.32\textheight,height=6cm]{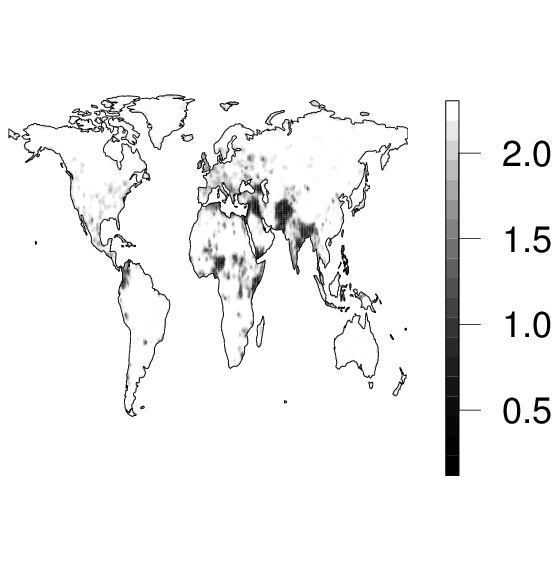} 
\vspace{-3.1em}
\caption{posterior standard deviation $\sigma_{\tilde{\zeta}(\bm{s},2013)}$}
  \label{fig:Pbisdrf12}
\end{subfigure}
\begin{subfigure}{0.32\textheight}
\vspace{0.0em}
\includegraphics[width=0.32\textheight,height=6.2cm]{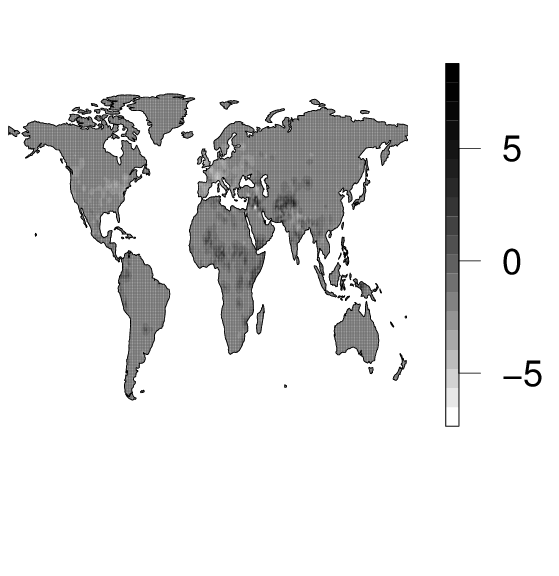} 
\vspace{-4em}
\caption{posterior mean log($\mu(\bm{s},2013)$)}
  \label{fig:Pprobsurf12}
\end{subfigure}
\vspace{1em}
\caption{Poisson model of the frequency of lethal terrorist attacks with GMRF posterior mean $\tilde{\zeta}(\bm{s},t)$ (\textit{left}), posterior standard deviation $\sigma_{\tilde{\zeta}(\bm{s},t)}$ (\textit{centre}), and posterior mean of the frequency of lethal attacks $\mu(\bm{s},t)$ (\textit{right}) (logarithmic scale) estimated in the 9,697 locations of mesh vertices and interpolated for all locations $\bm{s} \in \mathbb{S}^2$. Illustrative projected maps provide values in land surface for years $t=2002$ (\textit{top}) and $t=2013$ (\textit{bottom})} 
  \label{fig:Pprob}
\end{sidewaysfigure}

As a further step, we explore the spatial dynamics of both the lethality of terrorism and the frequency of lethal terrorist attacks by visualising relevant parameters that vary in both space and time. For this purpose, the posterior mean ($\tilde{\zeta}(\bm{s},t)$) and standard deviation ($\sigma_{\tilde{\zeta}}(\bm{s},t)$) of the GMRF provide valuable insight into the understanding of the spatial dynamics of terrorism and more particularly, of the uncertainty in the predictions of both the lethality of terrorism ($\pi(\bm{s},t)$) and the frequency of lethal terrorist attacks ($\mu(\bm{s},t)$). High values of $\sigma_{\tilde{\zeta}}(\bm{s},t)$ signify that there is high uncertainty with regard to the estimated values of $\tilde{\zeta}(\bm{s},t)$, mainly due to the scarcity or absence of data. Some areas have not encountered any terrorist attack during the entire period and therefore exhibit persistently high values, such as Siberia, the Amazonian region, Central Australia, or Greenland (figures~\ref{fig:Bbisdrf1}, \ref{fig:Bbisdrf12}, \ref{fig:Pbisdrf1}, \ref{fig:Pbisdrf12}). In contrast, several regions in South America, Africa, Golf Peninsula, India and Pakistan show an increase of the lethality of terrorism and the frequency of lethal terrorist attacks (figures~\ref{fig:Bprobsurf1}, \ref{fig:Bprobsurf12}, \ref{fig:Pprobsurf1}, \ref{fig:Pprobsurf12}) between 2002 and 2013 accompanied by lower values of $\sigma_{\tilde{\zeta}}(\bm{s},t)$. As expected, uncertainty is much lower in regions which encounter more attacks. 

\subsection{Hot-spots}
\label{subsec:hotspots}
\begin{sidewaysfigure}
\centering
\vspace{-5em}
\begin{subfigure}{0.48\textheight}
\includegraphics[width=0.48\textheight,height=8cm]{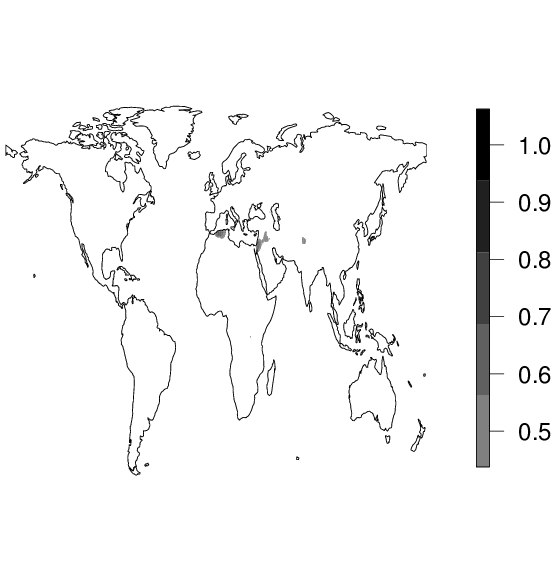}
\vspace{-4em}
\caption{Hot-spot of lethality in 2002 ($L_{CI_{95\%}} \pi(\bm{s},2002)>0.5$)}
  \label{fig:Bhot1}
\end{subfigure}
\begin{subfigure}{0.49\textheight}
\includegraphics[width=0.49\textheight,height=8cm]{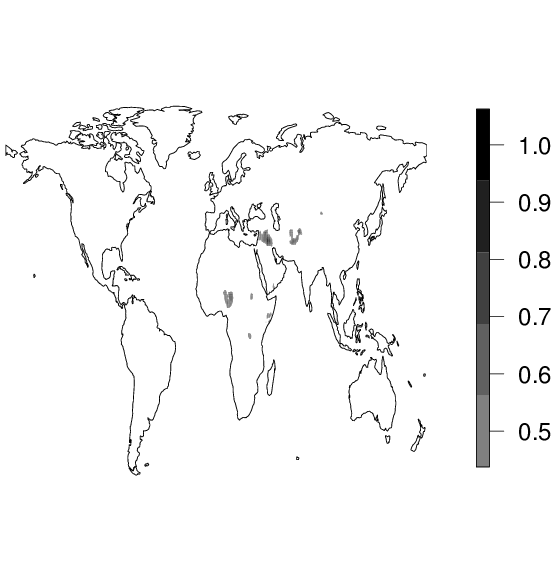} 
\vspace{-5.2em}
\caption{Hot-spot of lethality in 2013 ($L_{CI_{95\%}} \pi(\bm{s},2013)>0.5$)}
  \label{fig:Bhot12}
\end{subfigure}
\begin{subfigure}{0.49\textheight}
\vspace{-3em}
\includegraphics[width=0.488\textheight,height=8cm]{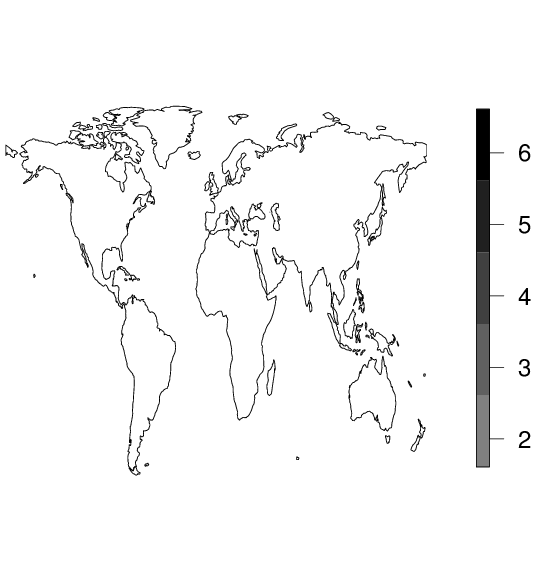}
\vspace{-4em}
\caption{Hot-spot of frequency in 2002 ($L_{CI_{95\%}} \mu(\bm{s},2002)>5$)}
  \label{fig:Phot1}
\end{subfigure}
\begin{subfigure}{0.49\textheight}
\vspace{-3em}
\includegraphics[width=0.488\textheight,height=8cm]{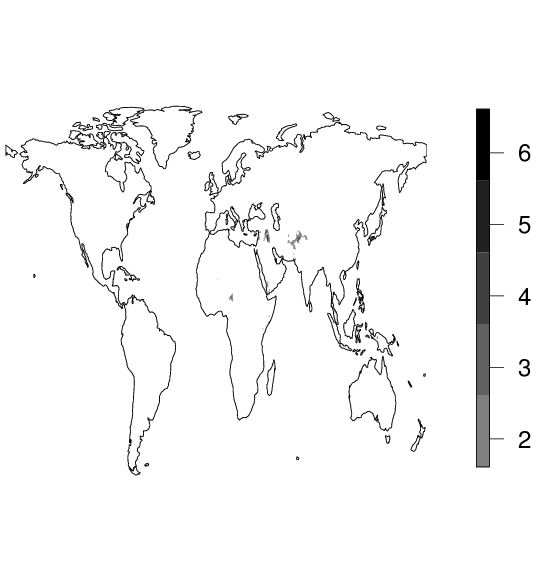} 
\vspace{-5.1em}
\caption{Hot-spot of frequency in 2013 ($L_{CI_{95\%}} \mu(\bm{s},2013)>5$)}
   \label{fig:Phot12}
\end{subfigure}
\caption[Hot-spots of lethality of terrorism and frequency of lethal terrorist attacks]{Hot-spots of lethality of terrorism across the world in 2002 (figure~\ref{fig:Bhot1}) and 2013 (figure~\ref{fig:Bhot12}), with lower bound of the 95\% credible intervals of posterior probabilities of lethal attack ($L_{CI_{95\%}}$) greater than 0.5. Hot-spots of frequency of lethal terrorist attacks across the world in 2002 (figure~\ref{fig:Phot1}) and 2013 (figure~\ref{fig:Phot12}), with $L_{CI_{95\%}}$ of posterior expected number of lethal attacks greater than 5. For illustrative purpose, we use a logarithmic scale ranging from 1.61 (log(5)) to 7.7 (log(2,153)). One may notice a general increase of terrorism activity in the Middle East and some African countries from 2002 to 2013 as illustrated by the presence of lethality (figure~\ref{fig:Bhot12}) and frequency (figure~\ref{fig:Phot12}) hot-spots in North-East Nigeria. }
\label{fig:hotspot}
\end{sidewaysfigure}

The identification of hot-spots of lethality and frequency of lethal attacks is highly valuable since it highlights areas more vulnerable to terrorism, which call for increased vigilance. For example, one might observe important changes in both the lethality of terrorism (figures~\ref{fig:Bhot1}, \ref{fig:Bhot12}) and frequency of lethal terrorist attacks (figures~\ref{fig:Phot1}, \ref{fig:Phot12}) in various locations in Iraq from 2002 to 2013. Both the number and the size of hot-spots in Iraq increased from 2002 to 2013, which reflects the intensification of terrorist activity that followed the invasion of Iraq in 2003 (Operation Iraqi Freedom) carried out by the coalition. In 2002 already, the CIA director George Tenet and the US National Intelligence Council warned the US and UK government that radicalization and terrorism activity will increase in Iraq and outside its borders due to the invasion of Iraq. This was later confirmed by Britain's Royal Institute of International Affairs (Chatham House) shortly after the 2005 London bombings, and also by various reports from Israeli think tank, Saudi and French Intelligence in particular \citep[pp.~18-21]{Chomsky2006}.

Similarly, we observe an increase in both the lethality of terrorism and frequency of lethal terrorist attacks in some regions in Afghanistan. Indeed, Afghanistan has been the theatre of numerous terrorist attacks since the US-led invasion in 2001 following 9/11. From 2003 to 2013, 3,539 terrorist attacks occurred mainly in the centre-east of Afghanistan, including the cities of Kabul, Jalalabad, Khogyani, and Sabari. From 2003 to 2013, 243 events (of which 157 were lethal) occurred in Kabul only. Most of these attacks were perpetrated by the Taliban. Even after the Taliban's withdrawal from Kabul in November 2001, lethal terrorist attacks did not cease in the city and within the country \citep{aljazeera2009}. Indeed, highly lethal suicide bombings intensified from 2006 to 2013 (and further) \citep{GTD2014}.

\subsection{Robustness tests}
\label{subsec:robustness}
The present study used the default Gaussian priors (multivariate Normal) provided by \texttt{R-INLA} for the parameters of the GMRF ($\kappa$ and $\tau$). As a robustness test, we run a prior sensitivity analysis for both the Bernoulli and the Poisson model changing the prior distribution of $\kappa$ and $\tau$. In practice, we changed the prior distribution of the variance of the GMRF $\sigma^2_{\tilde{\zeta}}$ and range $r$, whose quantities can be more easily interpreted, and therefore attributed a prior distribution. For both, the Binomial and the Poisson model, we set $\sigma^2_{\tilde{\zeta}}=50$, which corresponds to a relatively large variance of the GMRF, since we assume that the spatial structure might exhibit considerable variation among areas that encountered a high number of lethal terrorist attacks (e.g.\ some locations in Iraq, Pakistan or Afghanistan) and those that did encounter only a few or none (e.g.\ some locations in Portugal, Brazil or Alaska).

Assuming that the lethality of terrorist attacks can spread over relatively large areas (e.g.\ through demonstration and imitation processes promoted by the media \citep{Brosius1991, Enders1992, Brynjar2000}), we set $r=500$ [km] in the Binomial model. In contrast, we set $r=100$ [km] in the Poisson model, since we believe that the number of lethal attacks is very specific to the characteristics related to the close neighbourhood in which they occur. The number of potential high-value, symbolic targets, human and public targets (see Section~\ref{subsec:covariates}) can vary widely across distant areas. For example, one might reasonably assume that Baghdad shares important similarities with close Iraqi cities such as Abu-Grahib or Al-Fallujah (approximately 30-60 km from Baghdad). However, distant cities such as Al-Kasrah, Iraq (approximately 300 km from Baghdad) might exhibit important different characteristics, and therefore numbers of lethal attacks that are almost independent of the levels observed in Baghdad. Hence, we compute the corresponding $\kappa$ and $\tau$ \citep{Lindgren2015,Bivand2015b}, $\kappa=\frac{\sqrt{8}}{r}$ and $\tau=\frac{1}{\sqrt{4\pi}\kappa\sigma}$ \citep{Lindgren2011}. 
\vspace{0pt}
\begin{figure}[h!]
\centering
\includegraphics[scale=0.32]{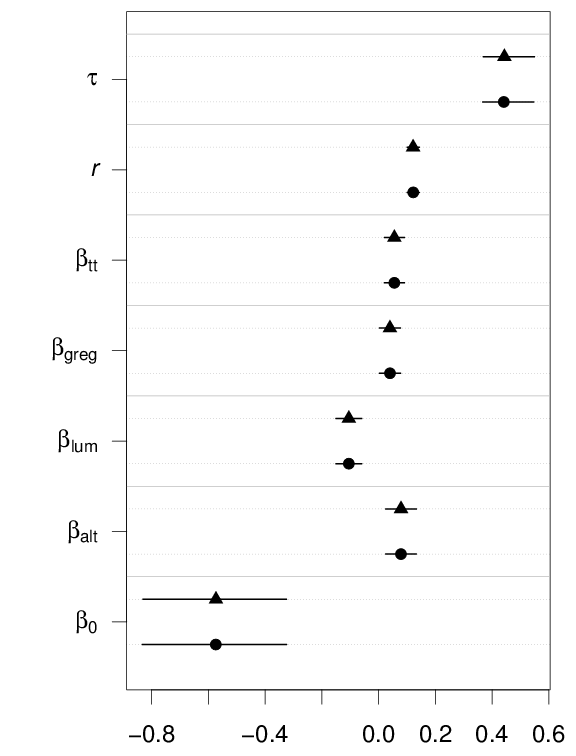}\hspace{1em}
\includegraphics[scale=0.32]{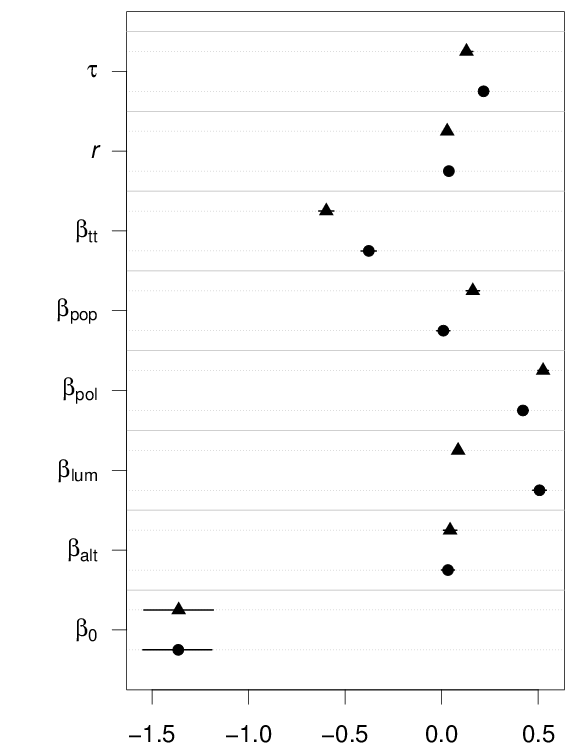}
\caption{Estimation of the mean and the 95\% credible intervals (line segment) of the posterior distribution of the parameters: intercept $\beta_0$, covariates $\bm{\beta}$, and spatial parameters (range $r$ and precision $\tau$ of the GMRF). The estimated values are provided for both the Bernoulli (\textit{left}) and the Poisson (\textit{right}) models. The mean of the posterior distribution of each parameter is illustrated for the default prior models (\mycircle{black}) and the models with the modified priors (\mytriangle{black}). The Bernoulli model uses modified priors set as: $\sigma^2_{\tilde{\zeta}}=50$ and $r=500$ [km], while the Poisson model uses $\sigma^2_{\tilde{\zeta}}=50$ and $r=100$ [km]. In both the Bernoulli and Poisson models, the results using the modified priors are consistent with those using the default priors in both point estimation (mean) and the direction of the effect}
\label{fig:priorcompar}
\end{figure}

The mean and the credible intervals of the estimated coefficients $\bm{\beta}$ and the parameters of the GMRF ($\tau$, $r$) are illustrated for both Bernoulli (Figure~\ref{fig:priorcompar}, \textit{left}) and Poisson models (Figure~\ref{fig:priorcompar}, \textit{right}), where default priors (\mycircle{black}) and modified priors (\mytriangle{black}) are specified. The parameters of the GMRF ($\tau$, $r$) are not affected by a change in prior in both Bernoulli and Poisson models, albeit the estimated value of the mean of $\tau$ is less robust in the Poisson model. In the Bernoulli model, both mean and credible intervals of all parameters are almost identical between the default and modified priors. In the Poisson models, the direction of the effect of the estimated coefficients is not affected by changes in prior, albeit the estimated values of the mean of the parameters (especially $\beta_{tt}$ and $\beta_{lum}$) are less robust. A higher sensitivity to change in prior is expected in the Poisson model since its number of observations ($n=6,386$) is considerably reduced compared to the Bernoulli model ($n=35,917$), thus the prior distribution has a higher influence on the posterior estimation, as illustrated in these results. 

\section{Discussion}
\label{sec:discussion}

This study proposes a Bayesian hierarchical framework to model both the lethality of terrorism and the frequency of lethal terrorist attacks across the world between 2002 and 2013. The statistical framework integrates spatial and temporal dependencies through a GMRF whose parameters have been estimated with \texttt{R-INLA}. The novelty of this study lies in its ability to systematically capture the effects of factors that explain the lethality of terrorism and the frequency of lethal terrorist attacks worldwide and at subnational levels. Moreover, the analysis of hot-spots at subnational level provides key insight into understanding the spatial dynamics of both lethality of terrorism and frequency of terrorist events that occurred worldwide from 2002 to 2013. In this Section, we highlight the main findings and limitations of this study, and suggest potential improvement, which could be carried out in future studies. 

Most country-level studies did not find significant linear relationship between the number of terrorist attacks and economic variables \citep{Krueger2003, Abadie2006, Drakos2006a, Krueger2008, Gassebner2011}. On a local scale, we showed that more economically developed areas tend to encounter more lethal attacks, which provides support to the theory advanced by \citet{Piazza2006}. The author suggested that more economically developed and literate societies with high standards of living may exhibit more lucrative targets, and therefore are expected to be more targeted by terrorist attacks. However, we also showed that terrorist attacks, when they occur, are less likely to be lethal in more economically developed areas. Despite that more economically developed areas are more prone to terrorism, they provide better protection to their targets, which reduces the risk of deadly attack. Therefore, one expect a smaller proportion of lethal attacks in more economically developed areas, as confirmed by our results. 

Similar to most country-level studies \citep{Abadie2006, Eubank2001, Li2005, Schmid1992}, we found that areas in democratic countries tend to encounter more lethal attacks compared to those in autocracies. The presence of freedom of speech, movement, and association in democratic countries might reduce the costs to conduct terrorist activities compared to those in autocratic countries \citep{Li2005}. In line with country-level \citep{Gassebner2011,Kurrild2006} and sub-national \citep{Nemeth2014} findings, we found that terrorist attacks are more likely to be lethal in ethnically diverse locations, perhaps due to stronger ethnic tensions \citep{Basuchoudhary2010}. As pointed out in \citet{Esteban2012}, one should acknowledge that ethnic diversity is only one possible measure of ethnic division among others, including \textit{ethnic fractionalization} and \textit{ethnic polarization}. Further analysis using different measures of ethnic division is required in order to assess the role of ethnic division in its wider sense.

While most country-level studies found significant positive linear relationship between population density and the number of terrorist attacks \citep{Ross1993, Savitch2001, Crenshaw1981, Swanstrom2002, Coaffee2010}, we did not find evidence that terrorist attacks are more likely to be lethal or that lethal attacks are more frequent in densely populated areas. However, the Euclidean distance from terrorist events to the nearest large city is positively associated with the lethality of terrorist attacks but negatively associated with the number of lethal attacks. Since targets are usually less secure in small cities and rural areas, this might facilitate deadly terrorist operations, which is consistent with our findings. In contrast, large cities offer greater anonymity and recruitment pool, which might open the door for a high number of lethal attacks.

In addition, more lethal attacks are expected within or close to large cities since they have an impact on a larger audience, which is often a desired outcome (\citealp[p.~41]{Laqueur1999}; \citealp[p.~115]{Crenshaw1990}; \citealp{Savitch2001}). Furthermore, terrorists benefit from high density communication network (road and rail) in large cities to move freely and rapidly from and to target points (\citealp{Heyman1980} ; \citealp[p.~189]{Wilkinson1979}). It is also not uncommon that terrorists target communication network infrastructure, as exemplified by the March 11, 2004 simultaneous attacks on several commuter trains in Madrid, Spain, which killed 191 people \citep{LAT2014}. 

Even though the number of lethal attacks itself does not appear to be associated with altitude, we found that terrorist attacks, when they occur, tend to be more lethal in higher altitude. Terrorists might be less constrained by governmental forces during terrorist operations launched in less accessible regions, such as mountains, which can provide save havens to terrorists \citep{Abadie2006, Ross1993}. Terrorist groups can therefore benefit from knowledge of ``rough'' terrain (mountainous regions) to defeat the enemy, as illustrated by the successful attacks carried out by the Mujahedeen groups against the Soviet Union, and later, the Taliban against NATO \citep{Buhaug2009}. Therefore, one might reasonably assume that terrorists increase their killing efficiency in mountains, which is reflected in a higher proportion of ``successful'' lethal attacks. However, since human targets are usually more scarce in mountainous regions, the advantage of the terrain might not suffice to compensate the lack of targets, which in turn reduces the number of possible lethal attacks that can be planned and carried out. Nevertheless, complementary analysis would be required in order to confirm the plausibility of this interpretation of the results. 

As with any statistical analyses of complex social phenomena, the outcome of this study should be taken with caution. First, since we aim to investigate terrorism across the entire world and at high spatial resolution, the availability of suitable covariates is limited. As a result, our study has ineluctably omitted numerous relevant drivers of both the lethality of terrorist attacks and their frequency, which include characteristics (e.g. psychological processes) of each member of terrorist groups, ideology, beliefs, and cultural factors (\citealp[p.~29]{Crenshaw1983}; \citealp[p.~151]{Wilkinson1990}; \citealp{Brynjar2000}; \citealp[pp.~92-93]{Richardson2006}), or reciprocal interactions between counterterrorism and terrorism \citep{English2010, Hoffman2002} for example. Despite the fact that our models do not allow for estimating the marginal effect of each potential unobserved factor, their aggregated effect has been however taken into account through the space-time dependence structure represented by the GMRF ($\tilde{\zeta}$ in equations~(\ref{eq:BHM2}) and (\ref{eq:PHM2})).

Second, one could reasonably expect some spatial variability in the lethality and the frequency of terrorist attacks, especially within large cities that are regularly targeted by terrorists. However, since terrorist events from GTD are reported at the centroid of the nearest city in which they occurred, spatial variability of terrorism's lethality and frequency within cities cannot be captured. Moreover, we assume that the spatial correlation in both the lethality of terrorism and the frequency of lethal terrorist attacks depends only on the distance between the locations of terrorist attacks (stationarity) and is invariant to rotation (isotropy). This assumption might be too restrictive, since it would be equally reasonable to assume that the spatial correlation related to mass-casualty attacks extends into a larger spatial range, via a broader diffusion through media for example. Further studies might investigate the use of non-stationary models, which are currently being developed for the model class that may be fitted with INLA \citep{Lindgren2011}.

Third, the temporal unit exhibits limitations as well, since the study period is discretised into 12 years (2002-2013), even though GTD provides day, month, and year for most events. Access to more computational power may allow further analysis to investigate variation in a monthly or weekly basis. Moreover, for computational reasons, we assume no interaction between spatial and temporal dependencies of the lethality and frequency of terrorist attacks, i.e. \textit{separable} space-time models, where the covariance structure can be written as the product of a purely spatial and a purely temporal covariance function for all space-time locations \citep{Gneiting2006}. In our models, $\tilde{\zeta}$ follows a simple autoregressive process (AR(1)) in time. In non-separable models, the dependencies structure in both space and time is usually highly complex \citep{Harvill2010}, and therefore more computationally demanding.

Fourth, subjective choices have been made throughout the entire modelling process, which might affect both the internal and external validity of our results. A major concern is the absence of consensus on the definition of terrorism \citep{Beck2013, Jackson2016}, and subjectivity is therefore inevitable \citep[p.~23]{Hoffman2006}. In line with \citet[pp. 24-25]{English2010}, we agree that is all the more important that studies on terrorism must clearly state how terrorism is understood. Accordingly, we use data from GTD, which clearly states the definition used to classify acts as terrorist events. Moreover, as with any Bayesian analysis, our study involves a degree of subjectivity with regard to the choice of priors. Because of our relatively large dataset, we are confident that the choice of priors does not influence our results, as confirmed by our prior sensitivity analysis (Section~\ref{sec:result}). However, subjectivity remains in the definition of the threshold for hot-spots. We have chosen ones which ensures probability of lethal attacks higher than non-lethal (Bernoulli) and expected number of lethal attacks that correspond to high percentile (Poisson). We recommend practitioners and researchers in the field of terrorism to take particular care in choosing cut-off values, which might vary according to the purpose of their study. 

Despite its aforementioned shortcomings, this study suggests a rigorous framework to investigating the spatial dynamics of the lethality of terrorism and the frequency of lethal terrorist attacks across the world and on a local scale. It assesses the uncertainty of the predictions, which is crucial for policy-makers to make informed decisions \citep[p.~64]{Zammit2013} or to evaluate the impact of counterterrorism policies \citep{Perl2007} for example. Ultimately, this research may provide complementary tools to enhance the efficacy of preventive counterterrorism policies. 

\clearpage
\small
\bibliographystyle{chicago}
\bibliography{references}
\end{document}